\documentclass[sigconf]{acmart}

\usepackage{float}

\title{RailEstate: \\
An Interactive System for Metro Linked Property Trends}
\author{Chen-Wei Chang}
\affiliation{
  \institution{Virginia Tech}
  \city{Alexandria}
  \state{VA}
  \country{USA}
}
\email{wilsonchang@vt.edu}

\author{Yu-Chieh Cheng}
\affiliation{
  \institution{Virginia Tech}
  \city{Alexandria}
  \state{VA}
  \country{USA}
}
\email{yj24@vt.edu}

\author{Yun-En Tsai}
\affiliation{
  \institution{Virginia Tech}
  \city{Alexandria}
  \state{VA}
  \country{USA}
}
\email{yunen@vt.edu}

\author{Fanglan Chen}
\affiliation{
  \institution{Virginia Tech}
  \city{Alexandria}
  \state{VA}
  \country{USA}
}
\email{fanglanc@vt.edu}

\author{Chang-Tien Lu}
\affiliation{
  \institution{Virginia Tech}
  \city{Alexandria}
  \state{VA}
  \country{USA}
}
\email{ctlu@vt.edu}

\renewcommand{\shortauthors}{Chang et al.}

\copyrightyear{2025}
\acmYear{2025}
\setcopyright{rightsretained}
\acmConference[SIGSPATIAL '25]{The 33rd ACM International Conference on Advances in Geographic Information Systems}{November 3--6, 2025}{Minneapolis, MN, USA}
\acmBooktitle{The 33rd ACM International Conference on Advances in Geographic Information Systems (SIGSPATIAL '25), November 3--6, 2025, Minneapolis, MN, USA}
\acmDOI{10.1145/3748636.3762799}
\acmISBN{979-8-4007-2086-4/2025/11}

\begin{document}
\begin{abstract}
Access to metro systems plays a critical role in shaping urban housing markets by enhancing neighborhood accessibility and driving property demand. We present RailEstate, a novel web-based system that integrates spatial analytics, natural language interfaces, and interactive forecasting to analyze how proximity to metro stations influences residential property prices in the Washington metropolitan area. Unlike static mapping tools or generic listing platforms, RailEstate combines 25 years of historical housing data with transit infrastructure to support low-latency geospatial queries, time-series visualizations, and predictive modeling. Users can interactively explore ZIP-code-level price patterns, investigate long-term trends, and forecast future housing values around any metro station. A key innovation is our natural language chatbot, which translates plain-English questions (e.g., “What is the highest price in Falls Church in the year 2000?”) into executable SQL over a spatial database. This unified and interactive platform empowers urban planners, investors, and residents to derive actionable insights from metro-linked housing data—without requiring technical expertise. A demonstration video of the system is available at \url{https://www.youtube.com/watch?v=ZLiz8S1UXsc}.

\end{abstract}



\maketitle

\begin{figure*}[t]
\centering
\includegraphics[width=0.75\textwidth]{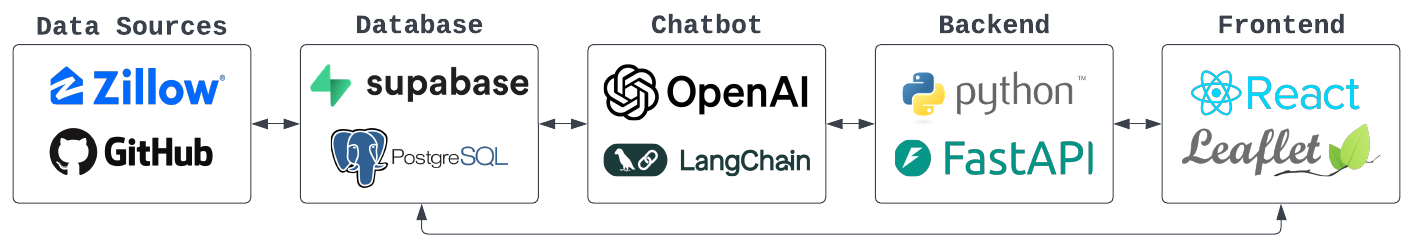}
\caption{System overview. Data pipeline with Supabase, OpenAI, FastAPI, and React.}
\label{fig:overview}
\end{figure*}

\section{Introduction}
\label{sec:intro}

Public transit accessibility is a well-established driver of urban real estate dynamics. Proximity to metro stations enhances neighborhood appeal by reducing commute times, increasing mobility, and improving connectivity~\cite{transit}. Studies consistently show a strong correlation between transit access and higher nearby property values~\cite{li2018impact, wen2018subway}. Yet homebuyers, investors, and planners still lack \emph{comprehensive, real-time} tools to reveal these effects in a localized, data-driven manner. Existing platforms offer only static listings without dynamic transit data or historical trends, leaving users to cross-reference maps, schedules, and market reports manually.

Prior work shows that transit proximity significantly affects property values, but tools for fine-grained, real-time spatial analysis remain limited. Statistical models quantify price effects~\cite{sun2016impact,shi2022does}, yet results vary across regions and time due to confounding factors~\cite{soltani2024nexus}. Spatial databases like PostGIS enable scalable queries~\cite{kothuri2002quadtree}, but are inaccessible to most non-technical users, while GIS dashboards often provide only static visualizations~\cite{kahn2007gentrification} without longitudinal or interactive analysis.

To address this gap, we present RailEstate, an interactive, database-driven web application that
delivers accessible, real-time, and interpretable insights into the relationship between metro access and residential housing trends. The system supports interactive geospatial queries around
metro stations and visualizes both current housing prices and long-term trends,
enabling users to observe fluctuations associated with infrastructure changes or economic
cycles. A natural-language chatbot further enhances users' access to complex
queries by translating plain-English questions into optimized SQL queries, which are executed over a cloud PostGIS backend.

RailEstate is a unified system that integrates spatial databases, interactive web mapping, and LLM-powered interfaces to support intuitive analysis of housing markets. Our contributions include:

\begin{itemize}
\item \textbf{Location-Aware Housing Price Engine.}
We deploy a cloud-hosted PostGIS database with spatial indexing to enable low-latency retrieval of housing prices near metro stations. Users can explore price gradients dynamically and derive actionable insights into transit-oriented value patterns.

\item \textbf{Interactive Map Visualization of Transit-Linked Prices}
Our React and Leaflet frontend visualizes ZIP-code-level average housing prices around selected metro stations, translating proximity analysis into an intuitive map-based interface, facilitating pattern recognition across urban subregions.

\item \textbf{Temporal Analytics Across Transit Zones.}
We curate and aggregate 25 years of Zillow housing data (2000–2025) to generate time-series visualizations of price trends surrounding each metro station. This enable users to analyze the long-term impact of transit developments (e.g., line extensions) on local housing markets.

\item \textbf{LLM-Powered NL2SQL Interface for Real-Time Data Retrieval}
By integrating LangChain and FastAPI, our system supports real-time translation of natural-language questions into executable SQL over the spatial database. Complex queries (e.g., “What is the highest price in Falls Church in the year 2000?”) are parsed, validated, and executed—democratizing access to geospatial insights for users without SQL expertise.
\end{itemize}

Together, these features form a responsive, AI-driven tool that delivers on-the-fly, transit-aware market intelligence—going beyond the capabilities of traditional mapping and real estate platforms.

\section{System Description}
In this section, we present the system architecture and core functions of RailEstate. We begin with an overview of the system architecture and workflow, then describe key functions that support interactive housing analysis.

\subsection{System Overview}
\label{sec:overview}

As shown in \autoref{fig:overview}, RailEstate uses a layered architecture with a React–Leaflet frontend, Supabase APIs with a FastAPI backend, a LangChain text-to-SQL engine (GPT-4o-mini), and a PostGIS PostgreSQL database containing Zillow prices, WMATA transit data, and ZIP code boundaries.

\textbf{Technology Stack.}
The system features a React and Leaflet frontend for interactive mapping, a FastAPI backend for API routing, and a LangChain module powered by GPT-4o-mini for natural language to SQL translation. Supabase hosts a PostgreSQL/PostGIS database with spatial indexing, secure access, and serverless deployment. We chose Supabase to host PostGIS for its auto-generated REST endpoints and built-in access controls, which reduce backend glue while keeping low-latency reads.

\textbf{Datasets.}
The system integrates three primary data sources: (i) Zillow monthly housing prices (2000–2025) for ZIP-code-level valuation trends~\cite{zillowhouse}, (ii) GTFS-based metro station and line data for spatial alignment~\cite{metro}, and (iii) ZIP code boundary GeoJSONs from GitHub for regional visualization and filtering~\cite{github}. While our demo targets the Washington Metrorail network, the pipeline is portable: any city with GTFS transit geometry and ZIP-level price time series can be onboarded without code changes to the query layer.

\textbf{Workflow.}
A typical workflow begins with selecting a metro station, which triggers spatial queries to retrieve nearby ZIP code prices and render overlays. Users may also submit natural language questions, translated by LangChain into SQL and executed on PostGIS, with results returned through the chatbot or map. This end-to-end pipeline runs in the browser and supports sub-second, interactive exploration of transit-aware housing trends.

\textbf{Implementation and Query Optimization.}
RailEstate’s implementation spans three tightly coupled components including \textbf{Interface \& Visualization}, \textbf{Backend \& Query Optimization}, and \textbf{Text‑to‑SQL Chatbot}, each engineered for interactive latency:

\begin{itemize}
    \item \textbf{Spatial and attribute indexing.} GiST/SP‑GiST indexes on geospatial columns (ZIP geometries, station buffers) and B‑tree indexes on frequently filtered attributes (state, value) reduce query latency by an order of magnitude.
    \item \textbf{Data cleaning and outlier management.} The ETL pipeline removes null records and caps extreme values at the 5th/95th percentiles to preserve statistical integrity.
    \item \textbf{Efficient joins and normalization.} Six normalized tables (\textit{Stations}, \textit{Lines}, \textit{Station\_Path}, \textit{Boundary}, \textit{Locations\_Prices}, and \textit{Predictions}) enable relational joins across spatial, temporal, and attribute dimensions with minimal redundancy.
    \item \textbf{Real‑time access.} Supabase’s auto‑generated RESTful APIs allow direct client queries without middleware overhead across 300k+ records.
\end{itemize}

\begin{figure}[t]
\centering
\includegraphics[width=0.88 \columnwidth]{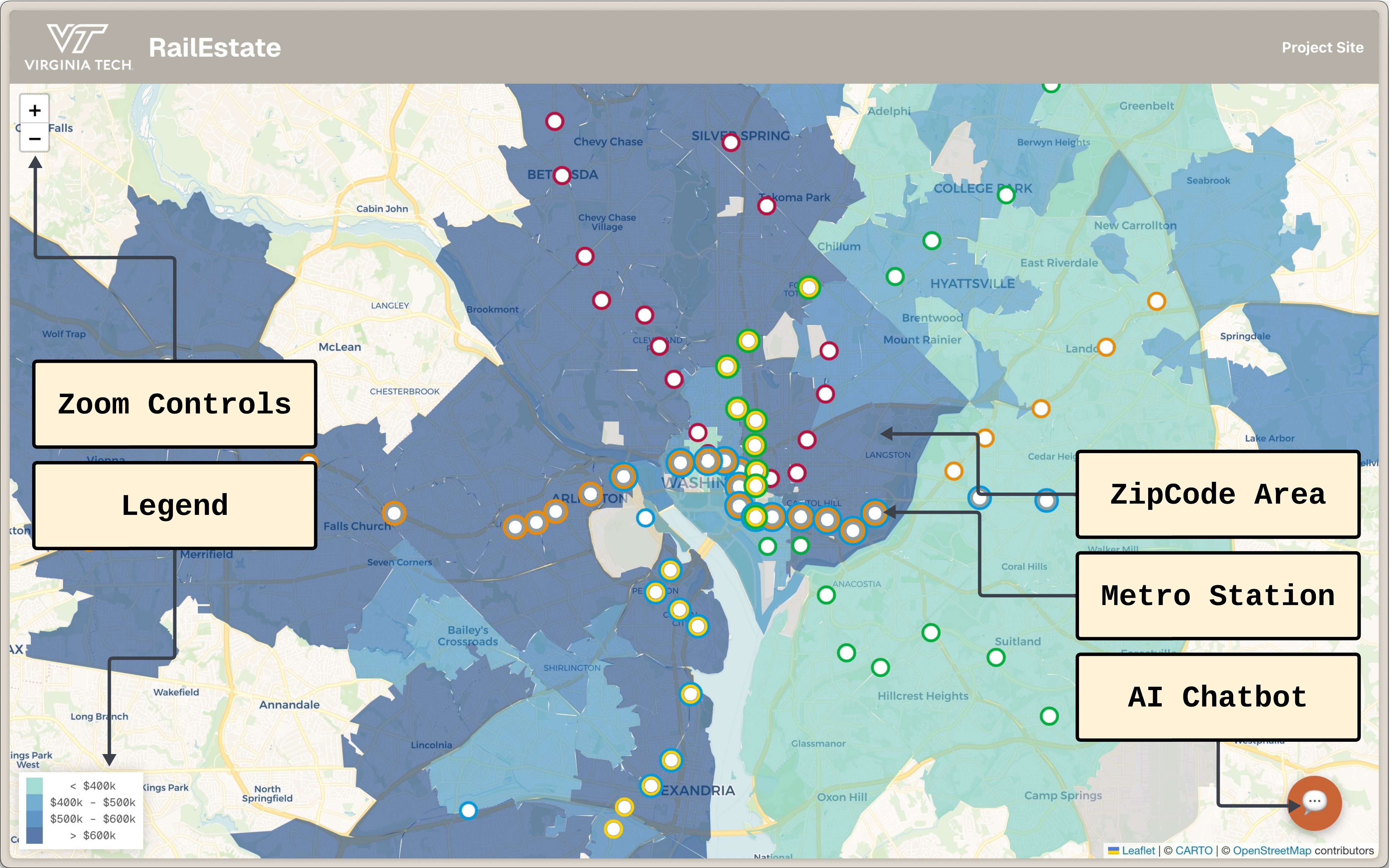}
\caption{Interactive map of metro stations and ZIP-code housing price bands. Basemap © CARTO}
\label{fig:metro}
\end{figure}

\begin{figure}[htbp]
\centering
\includegraphics[width=0.93\columnwidth]{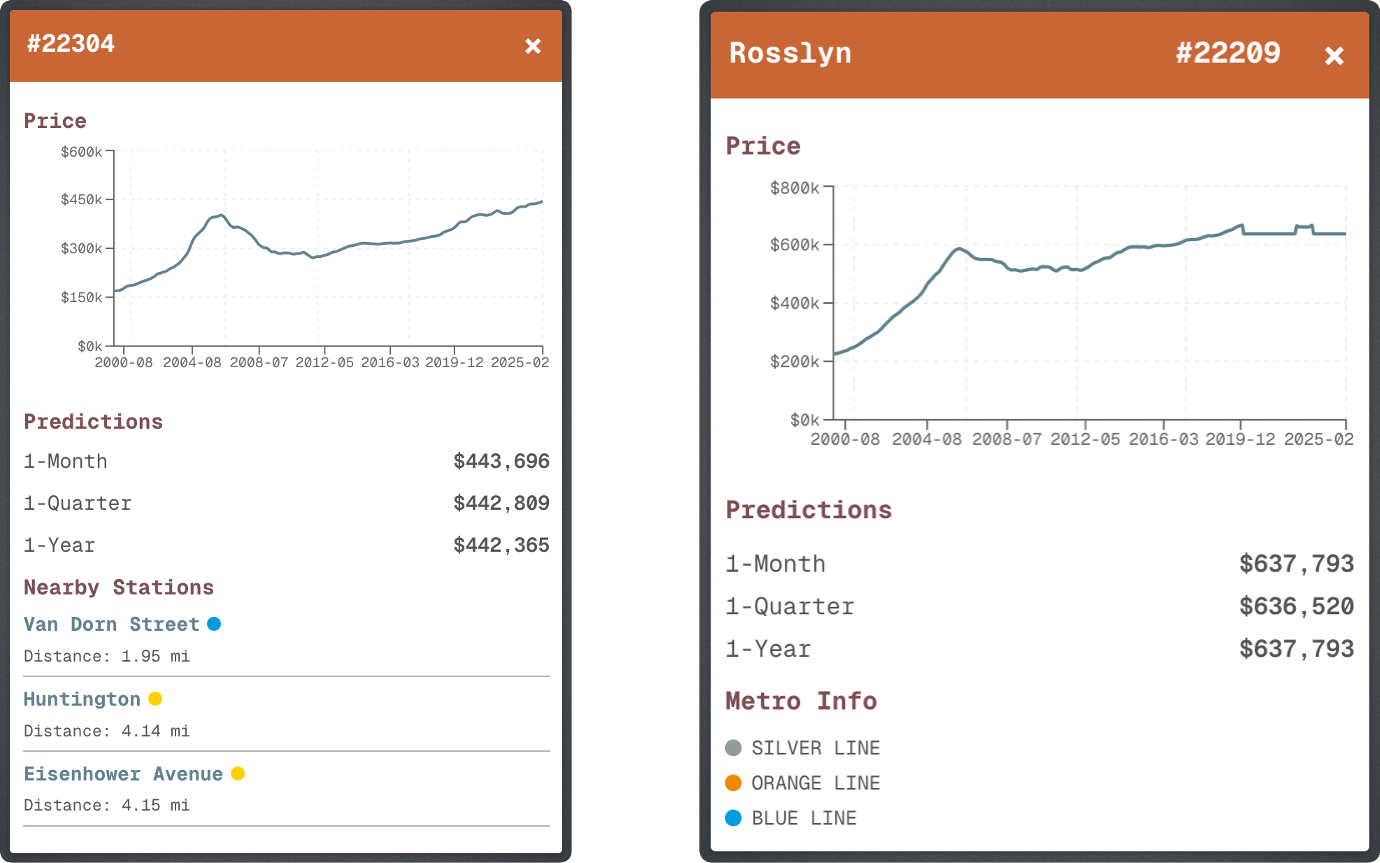}
\caption{Popups for ZIP code (left) and metro station (right) showing price trends and forecasts.}
\label{fig:popup}
\end{figure}

To demonstrate RailEstate’s capabilities, we outline its four core user-facing functions: location-aware price visualization, spatiotemporal trend analysis, housing price forecasting, and natural language querying.

\subsection{Function 1: Interactive Map for Location-Aware Price Visualization}
RailEstate’s interactive map (\autoref{fig:metro}) overlays metro stations and ZIP-code polygons on a CARTO basemap using Leaflet’s \texttt{GeoJSON} renderer. Regions are color-coded by ZIP-code-level average housing prices using fixed thresholds (e.g., $<400k, $400–500k, $500–600k, $>600k). These absolute bands, reflecting typical DMV price ranges, keep the legend stable across neighborhoods and months, avoiding the compression introduced by quantile schemes. Thresholds are configurable display parameters rather than affordability indicators. The interface further employs Leaflet’s layering system for polygons, station markers, and buffers, with smooth pan-and-zoom interaction. Core UI widgets—zoom controls, legend, AI chatbot, and ZIP highlights—are implemented as reusable React components (\textit{Header}, \textit{MapView}, \textit{PopupChart}, \textit{Legend}) to support maintainability and extensibility.

Click-activated pop-ups provide contextual details. ZIP code popups show the average price, trend summary, 25-year chart, ZHVF forecasts, and links to nearby stations (\autoref{fig:popup}, left), while station popups list the enclosing ZIP, historical/forecast charts, and served lines (\autoref{fig:popup}, right). Spatial matching relies on Turf.js operations (centroid distance, point-in-polygon, buffer tests) to ensure robust linkage between transit stops and housing regions.

\subsection{Function 2: Spatiotemporal Analysis of Housing Price Trends}
To reveal long‑term dynamics, RailEstate curates \textbf{25 years of Zillow monthly housing data} (2000–2025). For any chosen station and its surrounding ZIP code regions, the system renders responsive SVG time‑series plots (Recharts) that juxtapose price trajectories across regions. These charts capture broad economic cycles such as the 2008 housing crisis and local infrastructure shocks like the Silver Line extension, providing users with temporal insight into transit‑induced pricing patterns. Embedded directly within map pop‑ups, the charts rescale elegantly across devices, allowing users to switch interactively between absolute values and percentage growth without context loss.

\subsection{Function 3: Forecasting Future Housing Prices}
RailEstate augments historical analysis with forward-looking estimates based on \textit{Zillow Home Value Forecast} (ZHVF), which is a proprietary, AI-driven projection of the Zillow Home Value Index (ZHVI); the ZHVI itself is compiled from parcel-level valuations produced by Zillow’s Neural Zestimate model. Given an observed series, we recursively project one-, three-, and twelve-month horizons using

    \begin{equation}
        \text{ZHVI}_{t+1} = \text{ZHVI}_t \cdot \left(1 + \frac{\Delta_t}{100} \right),
    \end{equation}
    where $\Delta_t$ represents the percent change from month to month. The resulting values are stored in the \textit{Predictions} table and plotted beside historical curves so that users can compare recent performance with projected trajectories.

\subsection{Function 4: Natural Language Interface for Housing Queries}
To democratize access for non-technical audiences, RailEstate integrates a \textbf{Text‑to‑SQL chatbot} powered by FastAPI, LangChain, and GPT‑4o‑mini (\autoref{fig:chatbot}). When a user submits a question such as \textit{“What is the highest price in Falls Church in the year 2000?”} the backend executes the following pipeline:
\begin{enumerate}
    \item \textbf{Parse \& generate SQL.} LangChain’s \textit{SQLDatabaseChain} introspects schema metadata and emits semantically grounded SQL.
    \item \textbf{Execute.} The generated query runs directly on the Supabase PostGIS instance.
    \item \textbf{Post‑process.} A lightweight \textit{ChatOpenAI} chain formats the raw JSON into concise, context‑aware prose before returning it to the frontend.
\end{enumerate}
The choice of GPT-4o-mini balances strong language understanding with low latency, making it well suited for interactive use. It handles informal or ambiguous queries effectively, while a validation layer checks generated SQL for errors or unsafe operations to ensure robustness and security.

Backend helpers include an \texttt{ask(question)} routine for parsing, execution, and error handling; a \texttt{clean\_sql(sql)} filter for sanitization; and a \texttt{formatter} that converts tabular results into fluent English. On the client side, \textit{ChatbotAPI} manages asynchronous calls and \textit{ChatWidget} handles UI state and styling. Robust fallback logic further distinguishes out-of-scope queries (e.g., outside the DMV or unsupported years) and returns explanatory guidance, maintaining a high-quality conversational experience.

\subsection{Comparison with Existing Tools}
Unlike traditional real estate or transit dashboards, RailEstate integrates geospatial computation, forecasting, and NL2SQL capabilities in a unified system. Table~\ref{tab:comparison} highlights the key differences:
{
\renewcommand{\arraystretch}{1.5} 
\begin{table}[H]
  \centering
  \footnotesize
  \caption{Feature Comparison with Related Tools}
  \label{tab:comparison}
  \begin{tabular}{p{3.4cm}ccc}
    \toprule
    \textbf{Feature} & \textbf{Zillow} & \textbf{WMATA} & \textbf{RailEstate} \\
    \midrule
    Transit-Aware Spatial Queries   & $\times$ & \checkmark & \checkmark \\
    \hline
    Historical Price Visualization  & \checkmark & $\times$ & \checkmark \\
    \hline
    Forecasting Future Prices       & \checkmark & $\times$ & \checkmark \\
    \hline
    Natural Language Query (NL2SQL) & $\times$ & $\times$ & \checkmark \\
    \hline
    ZIP-to-Station Linking          & $\times$ & $\times$ & \checkmark \\
    \bottomrule
  \end{tabular}
\end{table}
}

Unlike Zillow and WMATA, RailEstate uniquely integrates transit-aware spatial queries, price forecasting, and NL2SQL in a single platform. It is also the only system that links ZIP codes to metro stations for localized, transit-informed analysis.

\section{Case Study: Real-Time NL2SQL Lookup for Historical Prices}
\label{subsec:case-fallschurch}

We present a case study to demonstrate the practical utility of RailEstate’s chatbot interface in exploring historical price trends.

\paragraph{Scenario.}
A user, interested in assessing the long-term investment potential, submits the following natural language query:  
\textit{“What is the highest price in Falls Church in the year 2000?”}

\paragraph{Workflow.}
The system processes the query through the following end-to-end steps:
\begin{enumerate}
  \item \textbf{Submit.} The query is entered in the \texttt{ChatWidget} and forwarded to the backend.
  
  \item \textbf{Generate \& Execute.} LangChain translates the user’s natural language query into SQL, then executes the generated statement on Supabase:
  
\begin{verbatim}
SELECT MAX(“value”) AS highest_price
FROM “Locations_Prices”
WHERE “city” = ‘Falls Church’ AND “date”
BETWEEN ‘2000-01-01’ AND ‘2000-12-31’;
\end{verbatim}
    This query retrieves the maximum housing price in Falls Church within the specified date range by filtering on city and applying an aggregate function.

  \item \textbf{Return.} The SQL is executed by the backend via LangChain over a secure Supabase PostGIS connection, returning the result as a structured JSON object.
  
  \item \textbf{Post-process.} A lightweight \texttt{ChatOpenAI} post-processing chain refines the raw JSON  into human-readable text.
  
  \item \textbf{Render.} The chat interface displays  \\
  \texttt{The highest price in Falls Church in the year 2000 was \$308,002.64}.
\end{enumerate}

\begin{figure}[t]
\centering
\includegraphics[width=0.99\columnwidth]{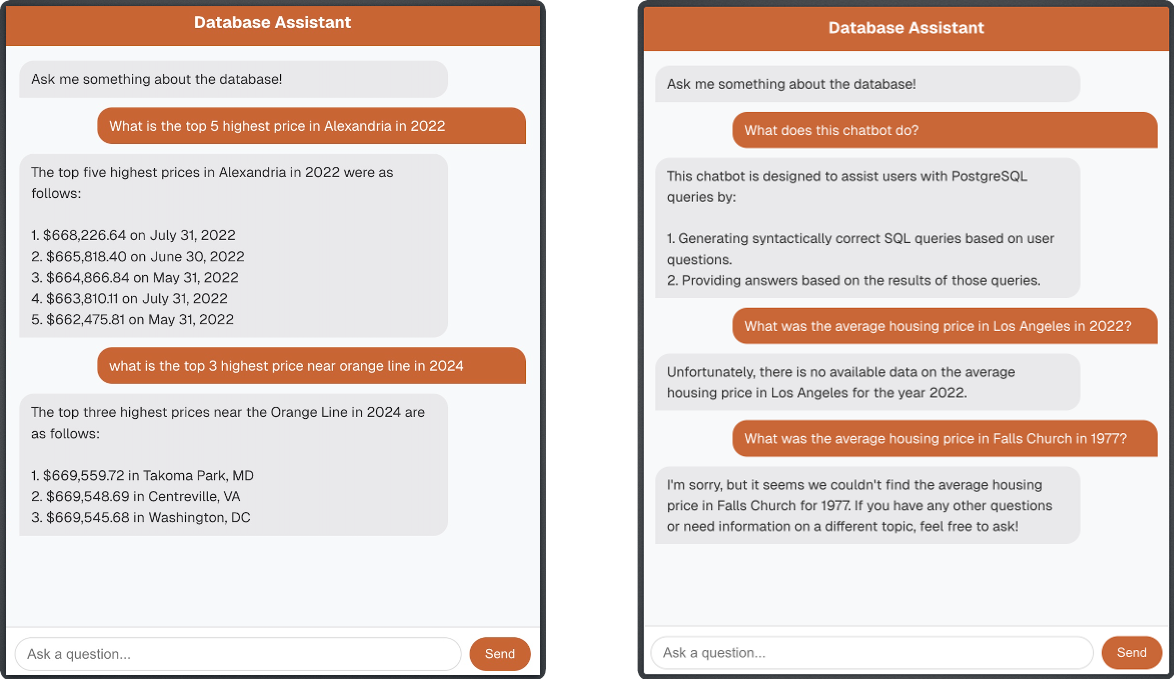}
\caption{Chatbot responses to valid queries (left) and unsupported queries (right).}
\label{fig:chatbot}
\end{figure}

\textbf{Robust Handling of Valid and Invalid Queries.}
To improve robustness, the chatbot distinguishes valid queries from unsupported ones. As shown in \autoref{fig:chatbot}, it returns structured answers for in-domain requests (left) and informative fallbacks for out-of-scope cases (right), ensuring reliability even with noisy input.

\section{Conclusion}

In this paper, we presented RailEstate, an interactive web-based system that integrates spatial databases, real estate data, and transit infrastructure to support metro-centric housing analysis in the Washington metropolitan area. Through real-time geospatial queries, historical trend visualization, and a natural language chatbot, RailEstate enables users to assess how metro proximity influences property values—without requiring technical expertise. Key features include an interactive map for visualizing price gradients, time-series charts of housing trends, forward-looking forecasts, and an NL2SQL chatbot for intuitive, SQL-free access to complex queries. Together, these components form a seamless and responsive platform that empowers urban planners, investors, and residents to make data-informed decisions.

\bibliographystyle{ACM-Reference-Format}
\bibliography{refs}

\end{document}